\documentclass[]{iopart}

\usepackage{iopams}  
\usepackage{graphicx,epsfig}
\begin{document}


\def\be{\begin{equation}}
\def\ee{\end{equation}}
\def\bea{\begin{eqnarray}}
\def\eea{\end{eqnarray}}
\def\degree{$^{\circ}$}

\title[]{
Lyapunov exponent and natural invariant density determination of chaotic maps: An iterative maximum entropy ansatz
}

\author{Parthapratim Biswas} 
\address{Department of Physics and Astronomy, The University of Southern Mississippi, MS 39406, USA} 
\ead{partha.biswas@usm.edu\footnote{Corresponding author}}

\author{Hironori Shimoyama} 
\address{Department of Physics and Astronomy, The University of Southern Mississippi, MS 39406, USA} 

\author{Lawrence R. Mead} 
\address{Department of Physics and Astronomy, The University of Southern Mississippi, MS 39406, USA} 

\begin{abstract} 

We apply the maximum entropy principle to construct the natural invariant density and Lyapunov 
exponent of one-dimensional chaotic maps.  Using a novel function reconstruction 
technique that is based on the solution of Hausdorff moment problem via maximizing Shannon 
entropy, we estimate the invariant density and the Lyapunov exponent of nonlinear maps in 
one-dimension from a knowledge of finite number of moments. The accuracy and the stability of the 
algorithm are illustrated by comparing our results to a number of nonlinear maps for which the 
exact analytical results are available. Furthermore, we also consider a very complex example for which 
no exact analytical result for invariant density is available.  A comparison of our results to those 
available in the literature is also discussed. 
\end{abstract}



\section{Introduction}

The classical moment problem (CMP) is an archetypal example of an inverse problem that involves reconstruction 
of a non-negative density distribution from a knowledge of (usually finite) moments~\cite{Hausdorff, Shohat, 
Akheizer,Mead,Tagliani, Abramov}.  The CMP is an important inverse problem that has attracted researchers from many diverse 
fields of science and engineering ranging from geological prospecting, computer tomography, medical imaging to 
transport in complex inhomogeneous media~\cite{inverse}. Much of the early developments in the fields 
such as continued fractions and orthogonal polynomials have been inspired by this problem~\cite{Shohat, Haydock}. 
The extent to which an unknown density function can be determined depends on the amount of information available in 
the form of moments provided that the underlying moment problem is solvable. 
For a finite number of moments, it is not possible to obtain the unique solution and one needs 
to supplement additional information to construct a suitable solution. The maximum entropy provides a suitable 
framework to reconstruct a {\it least biased} solution by simultaneously maximizing the entropy and satisfying 
the constraints defined by the moments~\cite{Jaynes}. 

In this communication we address how the maximum entropy (ME) principle can be applied to the Hausdorff moment 
problem~\cite{Hausdorff} in order to estimate the Lyapunov exponent and the associated natural invariant density 
of a nonlinear dynamical system. In particular, we wish to apply our ME ansatz to a number of nonlinear 
iterative maps in one-dimension for which the analytical results in the closed form are available. The problem was studied by 
Steeb \etal~\cite{Steeb} via entropy optimization for the tent and the logistic maps using the first few moments 
(up to 3). Recently, Ding and Mead~\cite{Ding1, Ding2} addressed the problem and applied their maximum entropy algorithm 
based on power moments to compute the Lyapunov exponents for a number of chaotic maps. The authors generated Lyapunov 
exponents using up to the first 12 moments, and obtained an accuracy of the order of 1\%. In this paper, we address 
the problem using a method based on an iterative construction of maximum entropy solution of the moment problem, and 
apply it to compute Lyapunov exponents and the natural invariant densities for a number of one-dimensional chaotic 
maps. Unlike the power moment problem that becomes ill-conditioned with the increasing number of moments, the 
hallmark of our method is to construct a stable algorithm by resort to moments of Chebyshev polynomials. The 
resulting algorithm is found to be very stable and accurate, and is capable of generating Lyapunov exponents 
with an error less than 1 part in $10^{3}$, which is significantly lower than any of the methods reported 
earlier~\cite{Steeb, Ding1}.  Furthermore, the method can reproduce the natural invariant density of the chaotic 
maps that shows point-wise convergence to the exact density function whenever available. 

The rest of the paper is organized as follows. In Section 2, we briefly introduce the Hausdorff moment problem 
and a discrete maximum entropy ansatz to construct the {\it least biased} solution that satisfies the moment constraints. 
This is followed by Section 3, where we introduce the natural invariant density as an eigenfunction of the 
Perron-Frobenius operator associated with the dynamical system represented by the iterative maps~\cite{Beck}. In Section 4, 
we discuss how the moments of the invariant density are computed numerically via time evolution of the dynamical variable, 
which are then used to construct the Lyapunov exponents and the natural invariant densities of the maps. Finally, in Section 5, we discuss the results of our 
method and compare our approximated results to the exact results and to those available in the literature. 

\section{Maximum Entropy approach to the Hausdorff moment problem}

The classical moment problem for a finite interval [a, b], also known as the Hausdorff moment problem, can be stated loosely 
as follows. Consider a set of moments
\be
\mu_i = \int_a^b x^i \, \rho(x)\, dx \quad \quad i = 0, 1, 2, \ldots, m, \quad i \le m
\label{eq-1} 
\ee
\noindent 
of a function $\rho(x)$ integrable over the interval with $\mu_i < \infty $  $\forall \, x \in $ [a,b] and 
$\rho(x)$ has bounded variation, the problem is to construct the non-negative function $\rho(x)$ from a knowledge 
of the moments.  The necessary and sufficient conditions for a solution to exist were given by 
Hausdorff~\cite{Hausdorff}. The moment problem and its variants have been discussed extensively in the 
literature~\cite{Shohat, Akheizer, Wimp, Junk} at length, and an authoritative treatment of the problem with 
applications to many physical systems was given by Mead and Papanicolaou~\cite{Mead}. For a finite number of 
moments, the problem is underdetermined and it is not possible to construct the unique solution from the moment 
sequence unless further assumptions about the function are made. Within the maximum entropy framework, one 
attempts to find a density $\rho_A(x)$ that maximizes the information entropy functional, 

\be 
S[\rho] = - \int_a^b \rho_A(x) \, \ln[\rho_A(x)] \, dx 
\label{eq-2} 
\ee

\noindent 
subject to the moment constraints defined by Eq.~(\ref{eq-1}). The resulting solution is an approximate 
density function $\rho_A(x)$ and can be written as
\be 
\rho_A(x) = \exp\left(-\sum_{i=0}^m \lambda_i \, x^i\right). 
\label{eq-3} 
\ee

The normalized density function $\rho (x)$ is often referred to as probability density by mapping the interval to [0,1] without 
any loss of generality.  For a normalized density with $\mu_0$ = 1, the Lagrange multiplier $\lambda_0$ is connected to the 
others via 
\bea
e^{\lambda_0} &=& \int_0^1 \exp\left(-\sum_{i=1}^m \lambda_i x^i \right) \nonumber 
\eea

A reliable scheme to match the moments numerically for the entropy optimization problem (EOP) was discussed by one of us in 
Ref.~\cite{AKB}. The essential idea behind the approach was to use a discretized form of entropy functional 
and the moment constraints using an accurate quadrature with a view to reduce the original constraint optimization 
problem in primal variables to an unconstrained convex optimization program involving dual variables. This guarantees the 
existence of the unique solution~\cite{note1}, which is {\it least biased} and satisfies the moment constraints defined 
by Eq.~(\ref{eq-1}). Using a suitable quadrature, the discretized entropy and the moment constraints can be expressed as respectively, 

\bea 
S[\rho] &=& -\int_0^1 \rho(x)\, \ln[\rho(x)]\, dx \, \approx \, - \sum_{j=1}^n \omega_j \, \rho_j \, \ln \rho_j  
\label{eq-4}\\ 
\mu_i &=& \int_0^1 x^i\, \rho(x)\, dx  \, \approx \, \sum_{j=1}^n\, (x_j)^i\, \omega_j \, \rho_j 
\label{eq-5} 
\eea 
\noindent 
where $\omega_i$'s are a set of weights associated with the quadrature and $\rho_j$ is 
the value of the distribution at $x=x_j$. If $\omega_j$ and $x_j$ are the weight and abscissas of the 
Gaussian-Legendre quadrature, the Eq.~(\ref{eq-4}) is exact for polynomials of order up to $2\,n-1$, 
and
\be 
\sum_{j=1}^n \omega_j = 1,  \qquad \qquad \sum_{j=1}^n \omega_j \, \rho_j = 1. 
\label{eq-6} 
\ee 
The task of our EOP can now be stated as, using $\tilde \rho_j = \omega_j \rho_j$ and $t_{ij} = (x_j)^i$, to 
optimize the Lagrangian 

\be 
L(\tilde \rho, {\bf \tilde \lambda}) = \sum_{j=1}^n \tilde \rho_j\, \ln \left(\frac{\tilde \rho_j}{\omega_j}\right) - \sum_{i=1}^m \tilde \lambda_i \left(\sum_{j=1}^n t_{ij} \, \tilde \rho_j - \mu_i\right) 
\label{eq-7} 
\ee 

where $ 0 \le \tilde \rho \in R^n$ and $\tilde \lambda \in R^m$, respectively are the primal and dual variables 
of the EOP, and the discrete solution is given by functional variation with respect to the unknown density, 
\be 
\tilde \rho_j = \omega_j \exp\left(\sum_{i=1}^m t_{ij} \, \tilde \lambda_i - 1\right), \quad j=1, 2, \ldots n. 
\label{eq-8} 
\ee 
The Eqs.~(\ref{eq-4}) to (\ref{eq-8}) can be combined together and a set of nonlinear equations can be 
constructed to solve for the Lagrange multipliers $\tilde \lambda$

\[
F_i({\bf \tilde \lambda}) =\sum_{j=1}^n t_{ij} \, \omega_j\, \exp \left(\sum_{k=1}^{m} t_{kj}\, \tilde \lambda_k - 1\right) 
- \mu_i = 0, \quad i = 1, 2, \ldots, m. 
\] 

The set of nonlinear equations above can be reduced to an unconstrained convex optimization problem involving 
the dual variables: 
\be 
\min_{\tilde \lambda \in R^m} \left[ D(\tilde \lambda) \equiv \sum_{j=1}^n \tilde \rho_j\,  \exp \left(\sum_{i=1}^m t_{ij} \, \tilde \lambda_i - 1\right) - \sum_{i=1}^m \mu_i \, \tilde \lambda_i \right].  
\ee 

By iteratively obtaining an estimate of $\tilde \lambda$,  $D({\tilde \lambda})$ can be minimized, and the EOP 
solution ${\rho}(\tilde \lambda^{*})$ can be constructed from Eq.~(\ref{eq-8}). In the equation above, 
$t_{ij} = x_j^i$ corresponds to power moments, but the algorithm can be implemented using Chebyshev polynomials 
as well. The details of the implementation of the above approach for shifted Chebyshev polynomials was 
discussed in Ref~\cite{AKB}. The maximum entropy solution in this case is still given by the Eq.(\ref{eq-3}) 
except that $x^i$ within the exponential term is now replaced by $T^{*}_i(x)$, where $T^*_i(x)$ is the shifted 
Chebyshev polynomials. In the following, we apply the algorithm based on the shifted Chebyshev  moments 
to construct the invariant density of the maps. 

\section{Lyapunov exponent and the natural invariant density of chaotic maps} 

The Lyapunov exponent of an ergodic map can be expressed in terms of the natural invariant 
density of the map: 
\be 
\Gamma = \int \rho(x) \, \ln |f^{\prime}(x)| \, dx 
\ee 
\noindent 
where $\rho(x)$ is the invariant density and $f^{\prime}(x)$ is the first derivative of the map $f(x)$ with respect 
to the dynamical variable $x$. The invariant density of a map can be defined as an eigenfunction of 
Perron-Frobenius operator associated with the map. Given an iterative map, $x_{n+1}$ = $f(x_n)$, one can construct 
an ensemble of initial iterates $\{x_0\}$ defined by a density function $\rho_0(x)$ in some subspace of the phase 
space and consider the time evolution of the density in the phase space instead of initial iterates $x_0$. The 
corresponding evolution operator $L$ is known as Perron-Frobenius operator, which is linear in nature as each 
member of the ensemble in the subspace 
evolves independently. The invariant density can be written as, 
\be 
L\, \rho(x) = \rho(x) 
\ee 
where $\rho(x)$ is a fixed point of the operator $L$ in the function space. In general, there may exist 
multiple fixed points but only one has a distinct physical meaning, which is referred to as the natural invariant 
density. Following Beck and Schl\"{o}gl~\cite{Beck}, the general form of the operator in one-dimension can be 
written as, 

\be
L\,\rho(y) = \sum_{x \in f^{-1}(y)} \frac{\rho(x)}{|f^{\prime}(x)|} \label{inv}. 
\ee

For an one-dimensional map, one can define the Lyapunov exponent as the exponential rate of 
divergence of two arbitrarily close initial points separated by $ \delta x_{n=0} = |x_0 - x^{\prime}_0|$ in the limit 
$n \to \infty$, and the exponent can be expressed as the average of the time series of 
the iterative map, 

\be
\Gamma = \frac{1}{N}\lim_{N \to \infty}\sum^{N-1}_{n=0}\ln |f'(x_n)|. 
\ee

For ergodic maps the time average of the Lyapunov exponent can be replaced by the 
ensemble average, 

\be
\Gamma = \int dx \; \rho(x) \; \ln|f'(x)|
\label{lamda}
\ee

\noindent 
using the natural invariant density. Equation (\ref{lamda}) suggests that the Lyapunov exponent can 
be obtained from a knowledge of the reconstructed natural invariant density from the moments. 
In the following we consider some nonlinear maps to illustrate how the normalized 
invariant density and Lyapunov exponent can be calculated using our discrete entropy optimization 
procedure.

\section{Reconstruction of invariant density as a maximum entropy problem} 
In the preceding sections, we have discussed how a probability density can be constructed from 
a knowledge of the moments (of the density) by maximizing the information entropy along with the 
moment constraints.  Once the density is reconstructed, the Lyapunov exponent can calculated from 
Eq.~(\ref{lamda}) using the reconstructed density. The calculation of the moments can proceed as 
follows. We consider a dynamical system represented by a nonlinear one-dimensional map, 

\be 
x_{n+1} = f(x_n) \nonumber \\
\ee 

where $n$=0, 1, 2, \ldots and $x_0 \in [0, 1]$.  The power moment of the time evolution of the 
iterate $x_n$ can be expressed as, 
\[
<x^{i}>  =  \lim_{t \to \infty} \frac{1}{t} \sum_{n = 0}^{t} (x_n)^i 
\] 

Since we are working with the shifted Chebyshev polynomials, the corresponding moments are, 
\be 
\mu_i =  \lim_{t \to \infty} \frac{1}{t} \sum_{n = 0}^{t} {T_i}^{*}(x_n) 
\label{moment}
\ee 

\noindent 
where $T_i^{*} (x) $ are the shifted Chebyshev polynomials and are related to Chebyshev polynomials via 
$T_i^{*}(x) = T_i(2\,x -1)$, and $ x \in [0, 1]$. 
A set of shifted Chebyshev moments can be constructed numerically from Eq.~(\ref{moment}), which 
can be used to obtain an approximate natural invariant density as discussed earlier. This 
approximate density can then be used to calculate Lyapunov exponents for the maps 
via Eq.~(\ref{lamda}).  By varying the number of moments, the convergence of the approximated invariant 
density function can be systematically studied and the accuracy of the Lyapunov exponent can be improved. 
We first apply our method to the maps for which the exact analytical results are available. 
Thereafter, we consider a nontrivial case where neither the Lyapunov exponent, nor the density can be 
obtained analytically and consists of a series of sharp peaks with fine structure which is difficult to 
represent using the form of analytical expression proposed by the maximum entropy solution.

\section{Results and Discussions}

Let us first consider the case for which the invariant density function and the Lyapunov exponent 
can be calculated analytically. We begin with the map, 

\be f_1(x)= \left\{ \begin{array}{ll}
                   \frac{2x}{1-x^2} \quad  & \mbox{$0\leq x \leq \sqrt{2}-1$} \\ \\ 
                   \frac{1-x^2}{2x} \quad & \mbox{$\sqrt{2}-1 \leq x \leq 1$}
                  \end{array} \right.  
\ee

\begin{figure}
\begin{center} 
\includegraphics[width = 5in, height=4.5in, angle = 0.0]{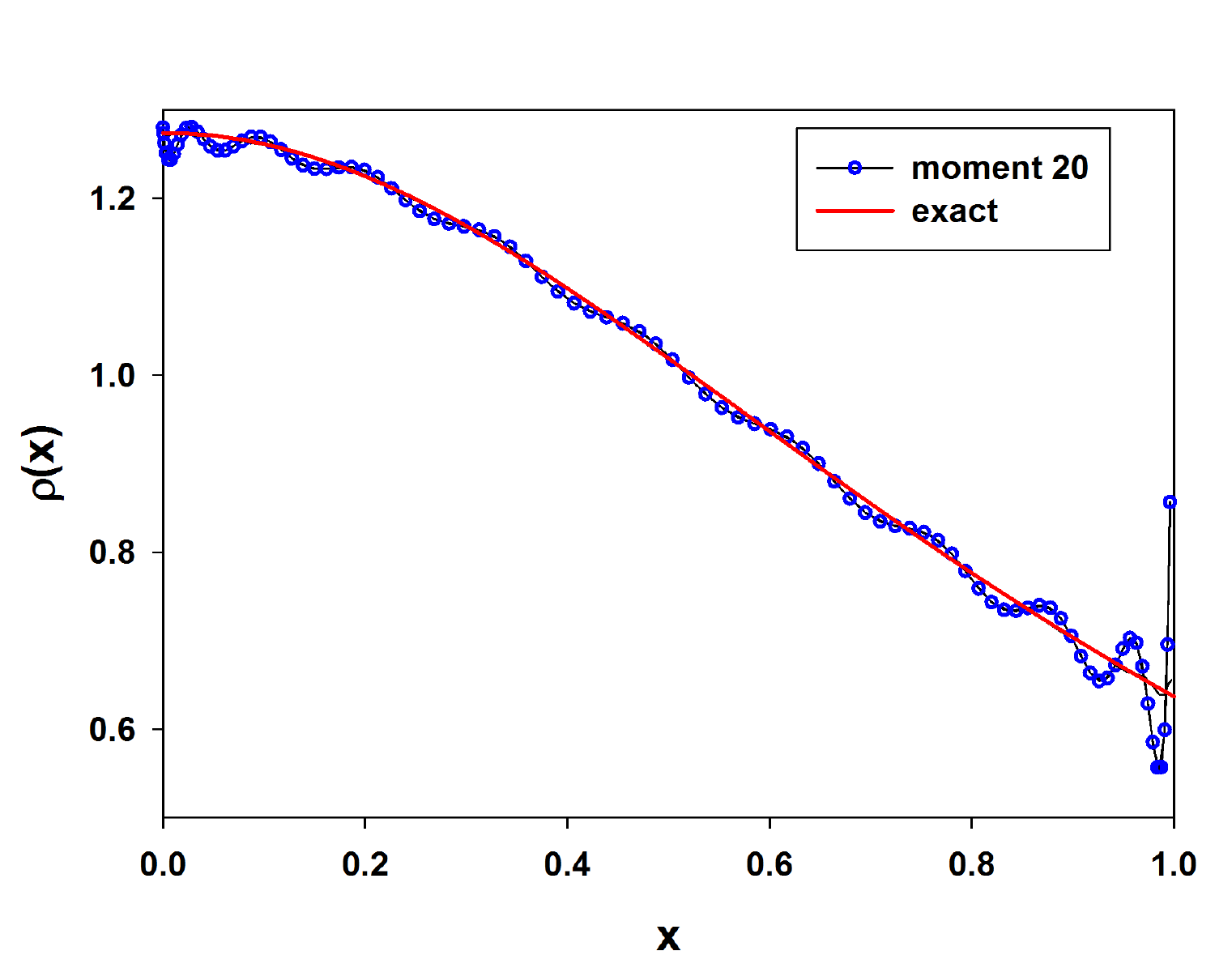}
\caption{ \label{fig1} 
The reconstructed density from the first 20 Chebyshev moments for the map $f_1(x)$ along with the 
exact density. Although the general shape of the density appears correctly, some oscillations 
are present in the data in absence of sufficient information. 
}
\end{center} 
\end{figure}

\begin{figure}
\begin{center} 
\includegraphics[width = 5in, height=4.5in, angle = 0.0]{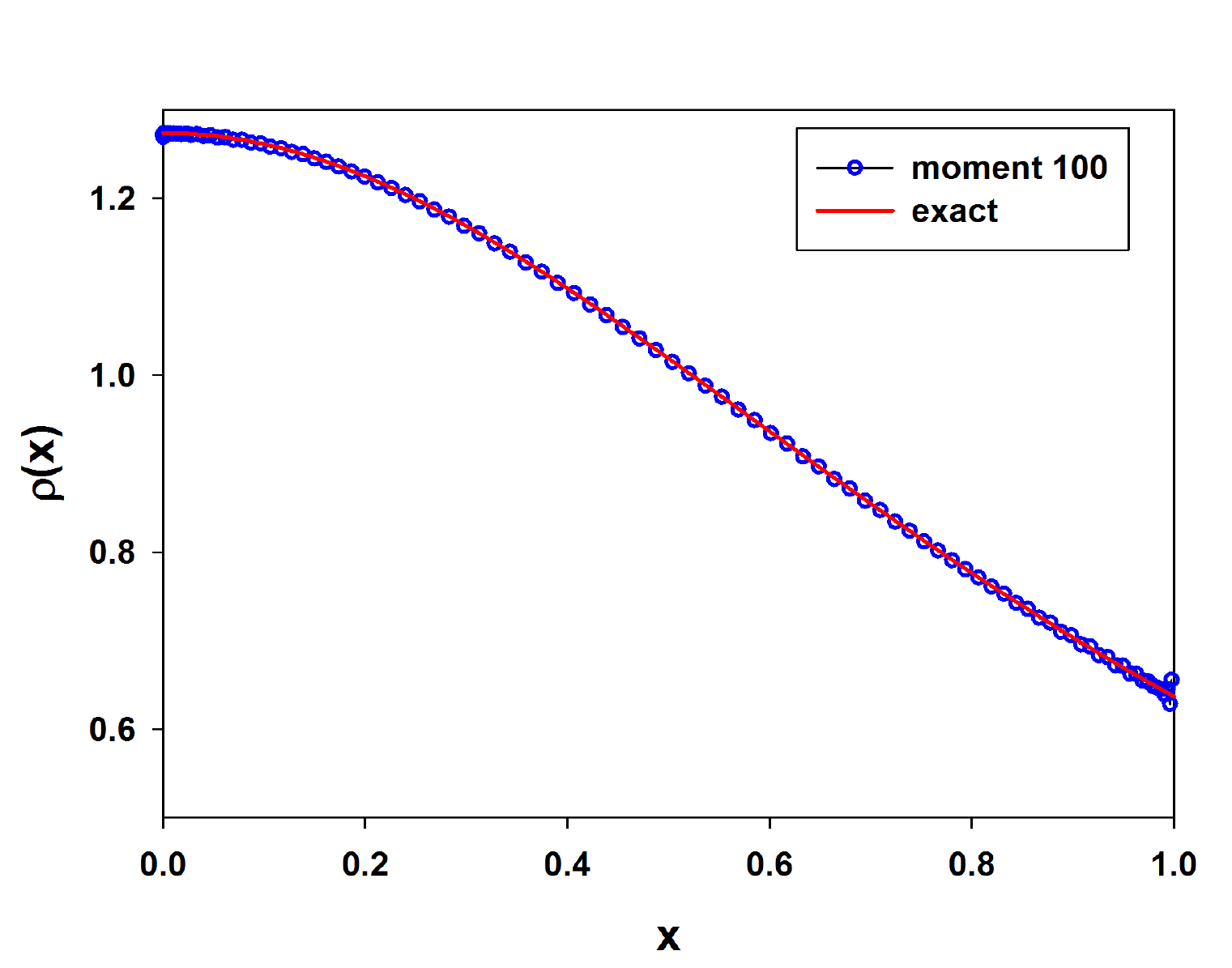}
\caption{ \label{fig2} 
The reconstructed density obtained from the first 100 moments for the map $f_1(x)$ along with the 
exact density. The approximate density now effectively matches point-by-point with the exact one 
with the exception of few points near the edges of the interval.
}
\end{center} 
\end{figure}

\noindent 
The invariant density for this map can be written as 
\be
\rho_1(x)=\frac{4}{\pi(1+x^2)} \label{f1}
\ee
and the Lyapunov exponent is given by $\ln 2$, which can be obtained analytically from Eq.~(\ref{lamda}). 
\begin{table}[htbp]
\caption{\label{tab1} Test Map $f_1$}
\begin{tabular}{lllcc}
Moments & $\Gamma_{\mathrm{maxent}}$ & Percentage error \\
\hline\hline
20 & 0.691577 & 0.226 \\
40 & 0.692786 & 0.055 \\
60 & 0.692999 & 0.021 \\
80 & 0.693061 & 0.012 \\
100 & 0.693109 & 0.006\\
\hline\hline
$\Gamma_{\mathrm{exact}}$ & $\ln 2 \approx$ 0.693147 \\
$\Gamma_{\mbox{{\scriptsize Ref.~\cite{Ding1}}}}$ & 0.69290  \\
\end{tabular}
\end{table}
The approximated Chebyshev moments for the map $f_1(x)$ can be obtained numerically 
from Eq.~(\ref{moment}). The ME ansatz is then applied to reconstruct the invariant 
density, and the Lyapunov exponent is obtained from this estimated invariant density.  
The results for the Lyapunov exponent are summarized in table \ref{tab1} for different set 
of moments.  The data clearly indicate that the approximated Lyapunov exponent rapidly converges to the 
exact value $\ln 2 \approx$ 0.693147 with the increase of number 
of moments. The error associated with the exponent is also tabulated, which shows that for the case 
of 100 moments the percentage error is as small as 0.006 reflecting the accurate and the stable 
nature of the algorithm.  In order to verify our method further, we now compare the approximated density to 
the exact density given by Eq.~(\ref{f1}).  This is particularly important because integrated 
quantities (such as Lyapunov exponent) are, in general, less sensitive to any approximation then the 
integrand (invariant density) itself, and that often makes it possible to get an accurate value 
of Lyapunov exponent from a reasonably correct density.  In figs.\ref{fig1} and \ref{fig2}, we 
have plotted the approximated densities for two different set of moments along with the exact density. 
Since the density is smooth and free from any fine structure, only the first 20 moments are found to be 
sufficient to get the correct shape of the density although some oscillations are present in the 
reconstructed density. 
On increasing the number of moments, the oscillations begin to disappear and for 100 moments the approximate 
density matches very closely with the exact one. The reconstructed density is shown in fig.\ref{fig2}, 
and it is evident from the figure that the density practically matches point-by-point with 
the exact density. 

As a further test of our method, we now consider the case of logistic map.  The map played a very important role 
in the development of the theory of nonlinear dynamical systems~\cite{log-map}, and can display a rather complex 
behavior depending on the control parameter $r$ defined via, 

\be
f_2(x)=r\; x\, (1-x). 
\ee

We consider three representative values of $r$ to illustrate our method in the chaotic and non-chaotic domain. 
In particular, we choose (a) $r=\frac{5}{2}$, (b) $r$=4, and (c) $r$ = 3.79285. The analytical densities are 
known only for the first two cases, and are given by respectively, 

\be 
\rho_2^{\mathrm{chaotic}}(x)=\frac{1}{\pi\sqrt{x-x^2}};  \quad \quad \, \, r=4 \\ 
\label{bucket1} 
\ee 
\be 
\rho_2^{\mathrm{fixed}}(x)=\delta\left(x-\frac{3}{5}\right);  \quad \quad r=\frac{5}{2}
\label{bucket2} 
\ee

For the remaining value of $r=3.79285$, no analytical expression for the density is known and the density 
consists of a number of sharp peaks along with some fine structure. The density in this case 
can be obtained numerically by iterating a set of initial $x_0$, and constructing a histogram
averaging over a number of configurations~\cite{Beck}. For the purpose of comparison to our maximum entropy 
results, we use this numerical density here. 

\begin{center}
\begin{table}[htbp]
\caption{\label{tab2}Logistic Map: Case A $f_2 = \frac{5}{2} x\,(1-x)$} 
\begin{tabular}{lllcc}
Moments & $\Gamma_{\mathrm{maxent}}$ & Percentage error \\
\hline\hline
10 & $-$0.693575 & 0.063  \\
20 & $-$0.693851 & 0.101  \\
30 & $-$0.693203 & 0.008  \\
40 & $-$0.693155 & 0.002 \\
\hline\hline
$\Gamma_{\mathrm{exact}}$ & $-\ln 2 \approx$ $-0.693147$ \\
\end{tabular}
\end{table}
\end{center}

\begin{center}
\begin{table}[htbp]
\caption{\label{tab3}Logistic Map: Case B $f_2=4\,x\,(1-x)$}
\begin{tabular}{lllcc}
Moments & $\Gamma_{\mathrm{maxent}}$ & Percentage error \\
\hline\hline
40 & 0.690703 & 0.35 \\
60 & 0.692101 & 0.15 \\
80 & 0.692850 & 0.04 \\
100 & 0.693319 & 0.02\\
\hline\hline
$\Gamma_{\mathrm{exact}}$ & $\ln 2$ $\approx$ 0.693147  \\
$\Gamma_{\mbox{{\scriptsize Ref.~\cite{Ding1}}}}$ & 0.68425  \\
\end{tabular}
\end{table}
\end{center}  

\begin{figure}
\begin{center} 
\includegraphics[width = 5in, height=4.5in, angle = 0.0]{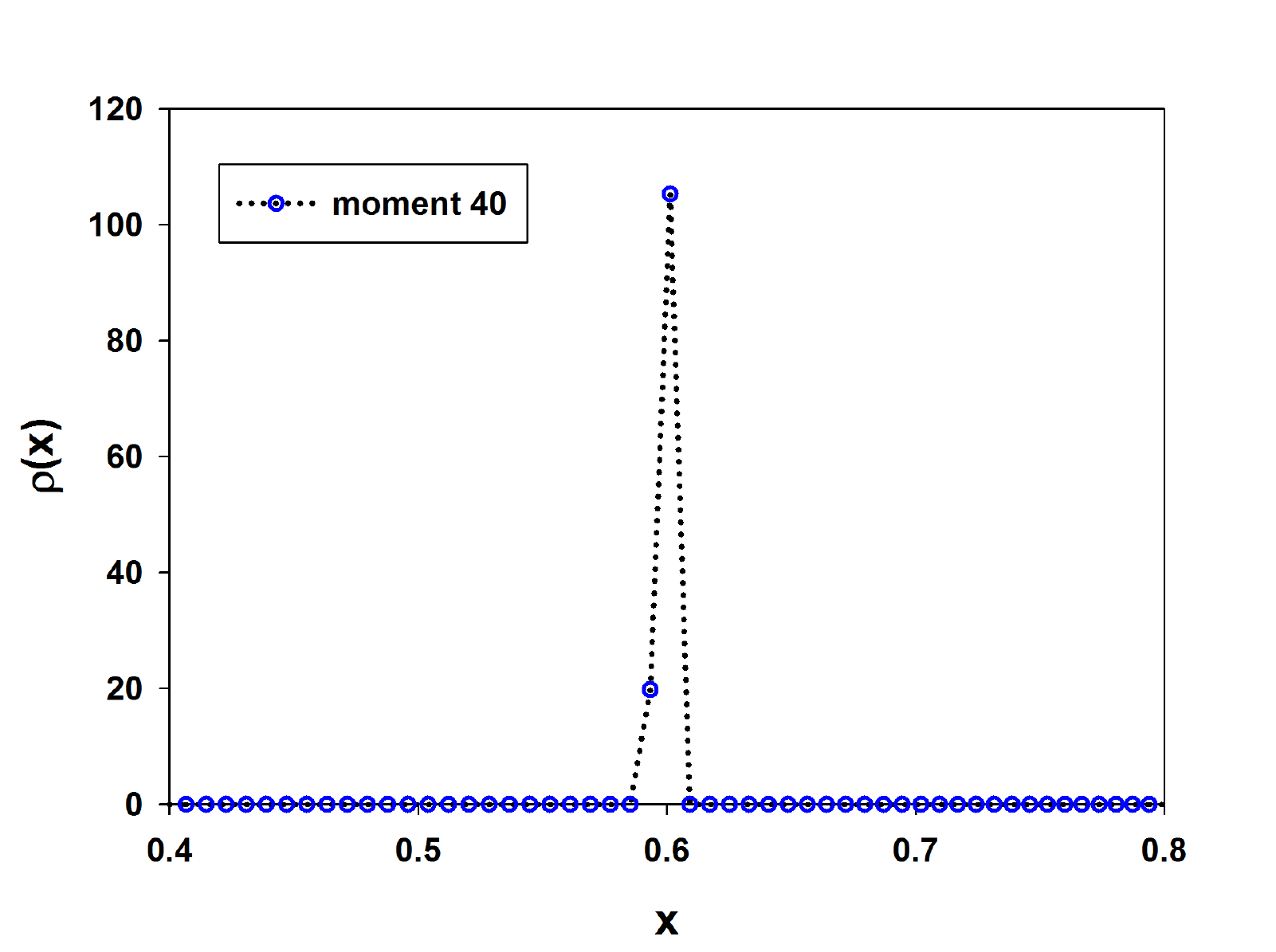} 
\caption{ \label{fig3} 
The reconstructed density for the logistic map for $r = \frac{5}{2}$ obtained from the first 
40 Chebyshev moments. The density consists of only two non-zero values in the 
interval [0.593, 0.601].  The exact density is a $\delta$-function centered at $x_0 = \frac{3}{5}$.
}
\end{center} 
\end{figure}

In tables \ref{tab2} and \ref{tab3} we have listed the values of the Lyapunov exponents for 
different number of moments for $r = \frac{5}{2}$ and $r = 4$ respectively. The errors associated 
with $\Gamma $ are also listed in the respective tables. The invariant density for $r = \frac{5}{2}$ is a 
$\delta$-function, and the exact analytical value of the exponent is given by $-\ln 2$. Since the 
invariant density is a $\delta$-function at $x_0=\frac{3}{5}$, it is practically impossible to 
reproduce the density very accurately using a finite number of quadrature points. However, our maximum entropy 
algorithm produces an impressive result by generating only two non-zero values in the interval 
containing the point $x_0=\frac{3}{5}$ using Gaussian quadrature with 192 points. The approximate 
density for a set of 40 moments is shown in fig.~\ref{fig3}. The two non-zero values of the density 
are given by 19.781 and 105.264 within the interval [0.593, 0.601]. It may be noted that for a 
normalized density, one can estimate the maximum height of the $\delta$-function to be of 
the order of $({\Delta x})^{-1} \approx 125.0$, where $\Delta x$ is the interval containing the 
point $x_0 = \frac{3}{5}$ point~\cite{note2}.  Furthermore, we have found that the result is almost 
independent of the number of moments (beyond the first 20), and the $\delta$-function has been observed to be correctly 
reproduced with few non-zero values using only as few as first 10 moments.  Table 3 clearly shows that 
the first 3 digits have been correctly reproduced using only the first 10 moments. On increasing 
the number of moments, there is but very little improvement of the accuracy of Lyapunov exponent. For each 
of the moment sets, the density is found to be zero throughout the interval except at few (two for the set 
40 and higher) points mentioned above. In absence of any structure in the density, higher moments do not 
contribute much to the density reconstruction, and hence it's more or less independent of the number of moments. Since the 
contribution to the Lyapunov exponent is coming only from the few (mostly two) non-zero values, and that these values 
fluctuate with varying moments, an oscillation of these values causes a mild oscillation in the 
Lyapunov exponent.

\begin{figure}
\begin{center} 
\includegraphics[width = 5in, height=4.5in, angle = 0.0]{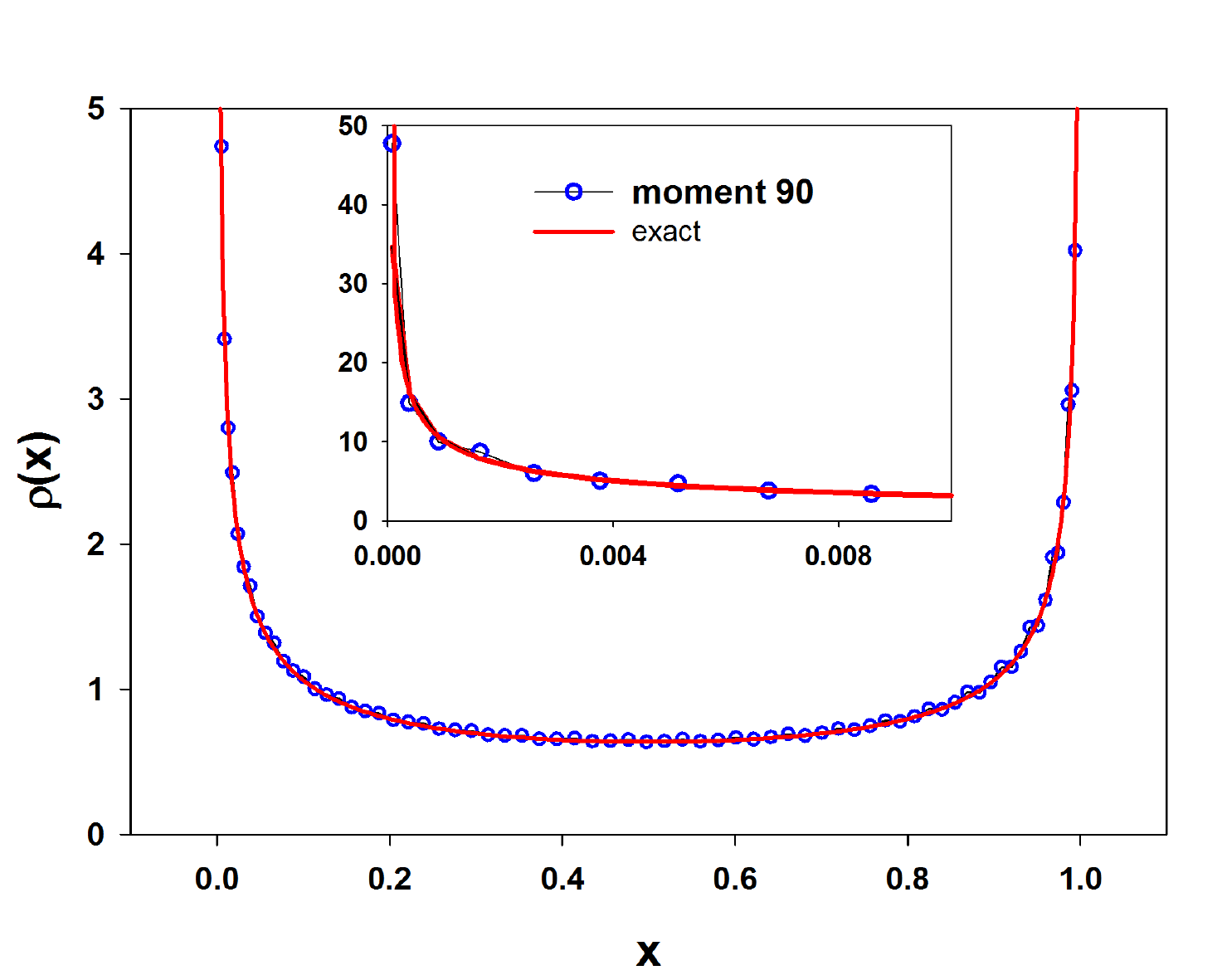}
\caption{\label{fig4} 
The reconstructed density for the map $f_2(x)$ for $r = 4$ obtained from the first 90 moments along with 
the exact density. The density matches excellently within the interval. The divergence behavior at the 
left edge near $x=0$ is also shown in the inset. 
}
\end{center} 
\end{figure}

We now consider the case $r = 4$. The exact density in this case is given by Eq.~(\ref{bucket1}) that has 
singularities at the end points $x$ = 0 and 1. It is therefore instructive to study the divergence behavior 
of the reconstructed density near the end points. In fig.\ref{fig4} we 
have plotted the approximate density obtained using the first 90 moments along with the exact density. The reconstructed 
density matches excellently within the interval. The divergence behavior near $x = 0$ is also plotted in 
the inset. Although there is some deviation from the exact density, the approximate density matches 
very good except at very small values of $x$. Such observation is also found to be true near $x = 1$. 
The results for the Lyapunov exponent are listed in table \ref{tab3} for different number of moments.  
It is remarkable that the exponent has been correctly produced up to 3 decimal points with 100 moments. 
While the error in this case is larger compared to the cases discussed before for the same number 
of moments, it is much smaller than the result reported earlier~\cite{Ding1}.  Our numerical 
investigation suggests that the integral converges slowly for Gaussian quadrature in this 
case owing to the presence of a logarithmic singularity in the integrand. This requires one to use more 
Gaussian points to evaluate the integral correctly. However, since the density itself has singularities at the end points, 
attempts to construct the density very close to the end points introduce error in the reconstructed density 
that affects the integral value. The use of Gauss-Chebyshev quadrature would ameliorate the latter problem, 
but for the purpose of generality (and in absence of prior knowledge of the density) we refrain ourselves from 
using the Gauss-Chebyshev quadrature. 

\begin{figure}
\begin{center} 
\includegraphics[width = 5in, height=4.5in, angle = 0.0]{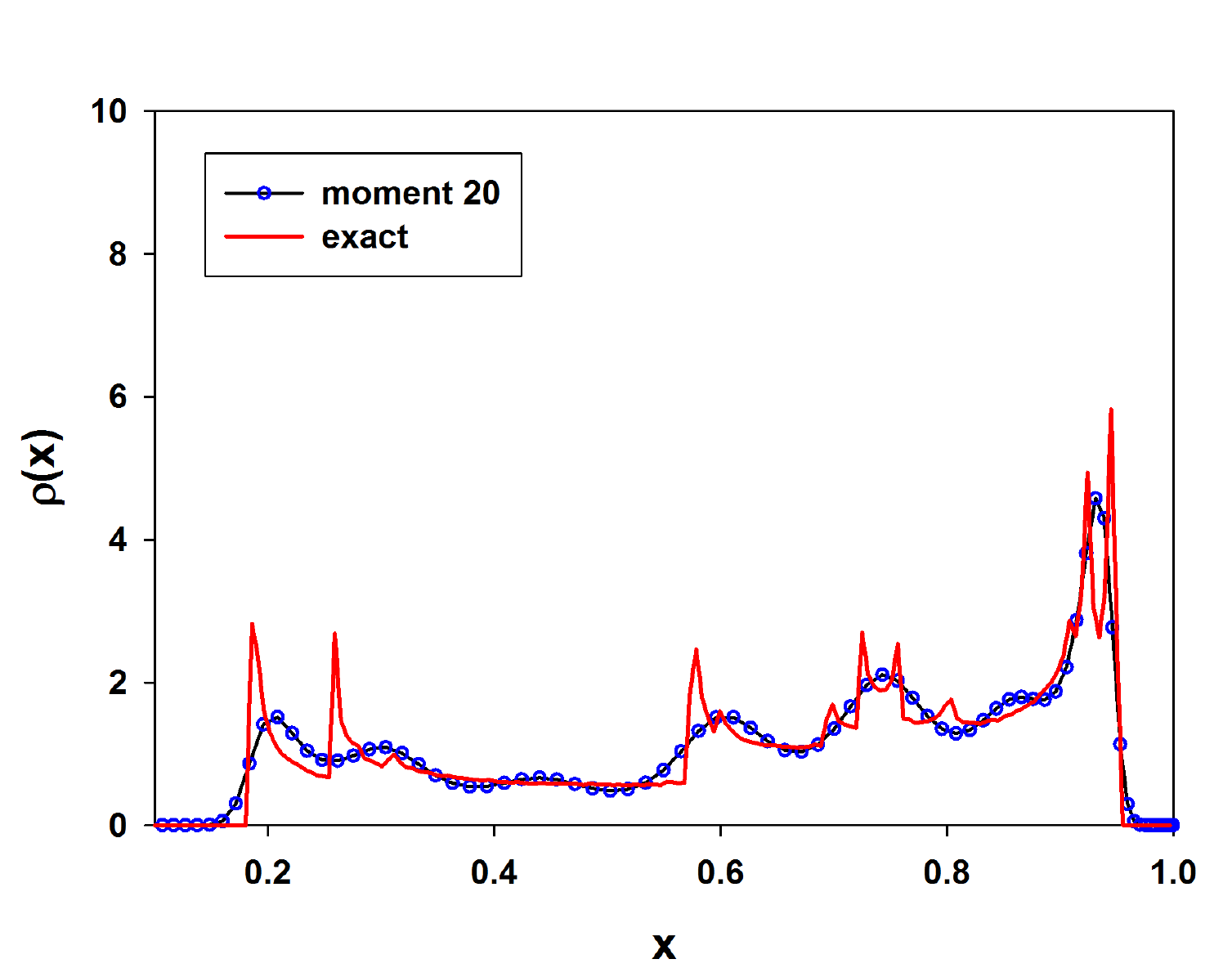}
\caption{ \label{fig5} 
The reconstructed density for the logistic map for $r = 3.7928$ from the 
first 20 Chebyshev moments along with the `exact' numerical density obtained via histogram method 
and averaged over 5000 configurations. While our maximum entropy algorithm produces most of peaks 
in the density, the fine structure of the peaks is missing in absence of information coming 
from the higher order moments. 
}
\end{center} 
\end{figure}

\begin{figure}
\begin{center} 
\includegraphics[width = 5in, height=4.5in, angle = 0.0]{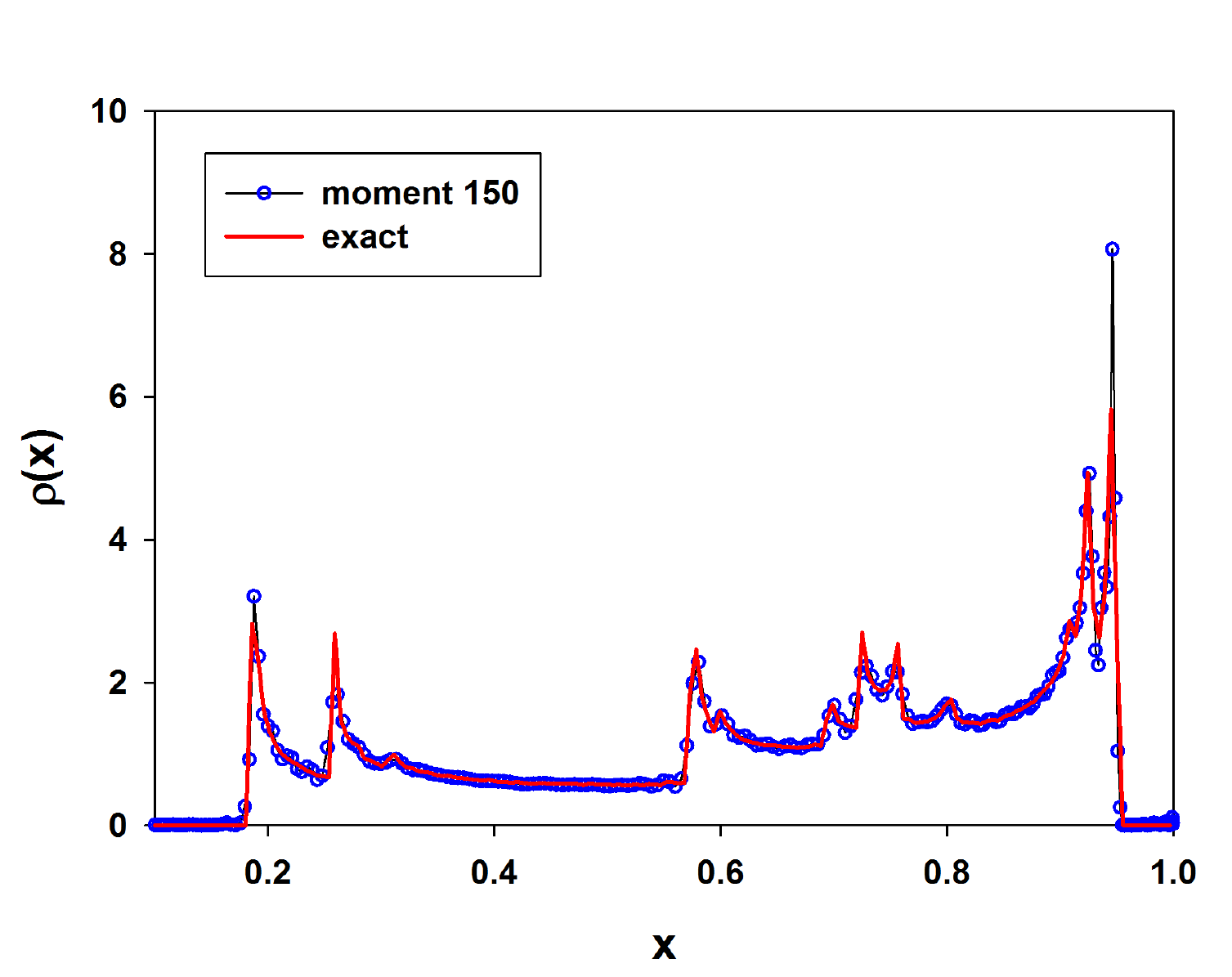}
\caption{ \label{fig6} 
The reconstructed density for logistic map for $r = 3.7928$ obtained from the first 150 Chebyshev 
moments. The corresponding `exact' density obtained from the map averaged over 5000 
configurations is  also plotted for comparison. The height and the position of the peaks are now 
correctly reproduced using the first 150 moments. 
}
\end{center} 
\end{figure}

Finally, we consider a case where analytical results are not available and the density consists of 
several sharp peaks having fine structure in the interval [0,1]. As mentioned earlier, the case 
$r= 3.79285$ for the logistic map provides such an example. The `exact' numerical density for this 
case is shown in fig.~\ref{fig5} along with the reconstructed density for 20 moments. The former is obtained by 
iterating several starting $x_0$ and constructing an histogram of the distribution of the iterates in 
the long time limit, which is then finally averaged over many configurations.  While our MEP ansatz produces most of the peaks in 
the exact density using the first 20 moments, the fine structure of the peaks is missing and so 
is the location of the peaks.  The reconstructed density can be improved systematically by increasing 
the number of moments, and for 150 moments the density matches very good with the exact density. In 
fig.~\ref{fig6} we have plotted both the reconstructed density for the first 150 moments 
and the numerical density from the histogram method. The result suggests that for sufficient 
number of moments our algorithm is capable of reproducing density which is highly irregular, 
non-differentiable and consists of several sharp peaks.

\section{Conclusion}

We apply an iterative maximum entropy optimization technique based on Chebyshev moments to calculate the 
invariant density and the Lyapunov exponent for a number of one-dimensional nonlinear maps.  The method consists of 
evaluating approximate moments of the invariant density from the time evolution of the dynamical variable of 
the iterative map, and to apply a novel function reconstruction technique via maximum entropy optimization 
subject to moment constraints. The computed Lyapunov exponents from the approximated natural invariant density are 
found to be in excellent agreement with the exact analytical values. We demonstrate that the accuracy of the 
Lyapunov exponent can be systematically improved by increasing the number of moments used in the (density) 
reconstruction process. An important aspect of our method is that it is very stable and accurate, and 
that it does not require the use of extended precision arithmetic for solving the moment problem. A comparison to 
the results obtained from power moments suggest that the algorithm based on Chebyshev polynomials 
gives more accurate results than the power moments. This can be explained by taking into account the superior minimax 
property of the Chebyshev polynomials and the form of the maximum entropy solution for Chebyshev 
moments~\cite{power, mason}. 
Our method is particularly suitable for maps for which exact analytical 
expression for invariant density are not available. Since the method can deal with a large number of moments, 
an accurate invariant density function can be constructed by studying the convergence behavior with respect to the 
number of moments. The Lyapunov exponent can be obtained from a knowledge of the invariant density of the maps. 
Finally, the method can also be adapted to solve non-linear differential and integral equations as discussed in 
Refs.\cite{Baker} and \cite{Mead2}, which we will address in a future communication. 
\vskip 0.1cm 

PB acknowledges the partial support from the Aubrey Keith Lucas and Ella Ginn Lucas Endowment in the form of a fellowship under 
faculty excellence in research program.  He also thanks Professor Arun K. Bhattacharya of the University of Burdwan 
(India) for several comments and discussions during the course of the work.

\begin{thebibliography}{99}
\bibitem{note1}
The solution is unique in the sense that it is {\it least biased} as far as entropy of the density 
is concerned, and the Hausdorff conditions are satisfied. There is no guarantee that the maximum entropy solution would be close to the exact 
solution particularly for very few moments. However, the quality of the MEP solution drastically 
improves with increasing number of moments, and numerical experiments for cases where exact solutions 
are known confirm that maximum entropy principle indeed can produce the correct solution.
\bibitem{note2} For a sufficiently small interval, $\delta (x - x_0)$ can be approximated by a box function 
of width $\Delta x$ around $x_0$ and of height $h$, giving $h \, \Delta x$ = 1 to satisfy the 
normalization condition. The height of the $\delta$-function obtained in our work is very close to this 
limiting value.
\bibitem{power} 
The presence of $\lambda_i x^i$ in the maximum entropy solution for power moments makes it very difficult 
to exploit information from the high order moments for $i > 20$ in the interval [0, 1] even with extended 
precision arithmetic. Such a problem does not appear in our formulation of the problem via Chebyshev moments 
owing to the bounded nature of Chebyshev polynomials, and consequently provides a way to incorporate 
systematically the information from the higher moments.
\end{thebibliography}
\end{document}